\newcommand{\beg}{\begin{equation}}
\newcommand{\en}{\end{equation}}
\def\nsf{{\rm {(-1)}^F NS}}
\documentclass{article}
\begin{document}
\thispagestyle{empty}
\hspace{11cm}hep-th/0408157
\vspace{2cm}
 \begin{center}
{\Large { Tachyon condensation at one loop} }\\

\vspace{0.7cm}
{\large { D. Nemeschansky\footnote{dennisn@physics.usc.edu} and V. Yasnov\footnote{yasnov@physics.usc.edu} }}

\vspace{0.3cm}
{\it Department of Physics and Astronomy \\University of Southern California\\Los Angeles, Ca 90089, USA}

\vspace{0.6cm}
\end{center}

\vspace{2cm}
\noindent
\begin{abstract}

Using  boundary string field theory we study the decay of unstable D-branes to lower dimensional D-branes via the tachyon condensation at one loop level. 
We  analyze one loop divergences and use the  Fischler-Susskind mechanism to cancel divergences arising at the boundary of moduli space. The tachyon action up to the second derivative is obtained and a logarithmic correction to the tachyon potential is written down explicitly. 
Multiple D-branes is also  considered and the role of the boundary fermions is highlighted. 
\end{abstract}
\newpage
\setcounter{page}{1}

\section{Introduction}
Tachyon condensation on D-branes has drawn  a lot of  attention during the last  few years. It is  essentially an off-shell process. Tachyon condensation can be described using 
the cubic string field theory \cite{witten1} or  the boundary string field theory (BSFT)\cite{witten2}. Although progress has been made in understanding  the 
decay of unstable D-branes to lower dimensional D-branes there  remains  many issues that need to be answered. Both the cubic string field theory and BSFT have been formulated on the disk. The 
 off shell actions have  been derived only in  tree level approximations. It is interesting to study  how the theories change when  string loop corrections are taken 
into account. The first loop  correction  is given by a boundary string field theory  on an annulus.

We will use BSFT to describe tachyon condensation. For  a quadratic boundary 
interaction the bosonic BSFT has been studied in \cite{moore0},\cite{gerasimov}. The exact tachyon potential has been  calculated along with value  of the  \mbox{D-b}rane tension.
The generalization to  superstrings was considered in \cite{moore}. It turns out that the supersymmetric case is   simpler than then bosonic one. The action is proportional
to the partition function on  the disk. This statement was rigourously proven using Batalin-Vilkovisky formalism in \cite{Niarchos}.

To include loop corrections one has to generalize the analysis to the annulus. BSFT  on an annulus has been considered by several authors \cite{andreev2},\cite{wung},\cite{berkeley},\cite{all},\cite{ar},\cite{andreev}.
Different authors have chosen different boundary conditions on the boundaries of the annulus. 
We will considered the case where the boundary conditions on the two boundaries are the same.
We argue that other boundary conditions lead to inconsistencies. In this paper we also study the Fischler-Susskind (FS) mechanism in the presence of the tachyon profile. This allows us to cancel divergences in the partition function that arise at the boundary of moduli space.

In the view of the recent development in cosmology \cite{gibbons} connected with Sen's proposal of the rolling tachyon \cite{sen}
it is also interesting to look how the string loop corrections modify the tachyon potential.

For the sake of completeness we want to mention that tachyon condensation was first introduced long time ago 
in the series of papers\cite{Bardakci}.

Before we start  with  our calculations let us summarize the assumptions  we will  make.  When all the  $\beta$-functions are linear the action on a the disk has the form  \cite{witten2}
\beg
\label{partition}
S=\left (1+\beta_i\frac{\partial}{\partial g_i}\right )Z(g_i),
\en
where $g_i$'s are  coupling constants and $Z(g_i)$ is the partition function on the  the disk. For superstrings the tachyon $\beta$-function is zero and Eq. (\ref{partition}) reduces  to \cite{moore}
\beg
S=Z.
\en       
We assume that this relation is still true on  the annulus. The new feature in BSFT on the annulus is that we encounter divergences from integration over moduli. The corresponding two dimensional
field theory must be renormalizable \cite{witten2}.
We will show that for the quadratic tachyon profile this is the case.  To cancel the infinities one has to introduce a counterterm. It  contributes
to the graviton $\beta$-function and depends on the tachyon coupling. The boundary RG is  mixed with closed string excitations and therefore one also needs the  closed string field 
theory. We assume that at least for the quadratic boundary interaction it is possible to restrict  to BSFT and ignore closed string field theory.

\section{Bosonic strings}
The boundary interaction for a bosonic  string on a disk $D$ has  the form
\beg
S_{bndry}=\frac{1}{2\pi}\int_{\partial D}\hspace{-2mm}d\theta\sqrt{h} T(X),
\en
where  $T(x)=uX^2$ is the  quadratic tachyon profile and $\sqrt{h}d\theta$ is the  length element on the boundary. The tachyonic action can be determined using  BSFT.  The tree level contribution is obtained from  a disk amplitude.
To include loop corrections one has to consider world sheets  with higher Euler characteristic. In this paper we consider the next term in the loop expansion. It 
corresponds to   world sheet with Euler characteristic $\kappa =0$. This surface has two  boundaries and it can  be  represented  either as a cylinder or as an annulus. For a  conformaly invariant theory  it does not matter which surface one  uses. However, a non-trivial  tachyon  profile breaks conformal invariance and the  partition functions on the annulus and cylinder  do not coincide. 

We will start our calculation by comparing the partition function on the annulus to the one on the disk. 
The boundary  of an annulus is 
 two circles, one  with the radius $1$, and  another one  with radius $a<1$. The choice of 
the tachyon profile on the outer boundary is fixed by the disk BSFT. On the other hand it is not obvious how to fix the tachyon profile on the inner boundary. It is clear that we want a tachyon profile that is 
 quadratic, but this leaves the freedom of having a different normalization for the quadratic term on the two boundaries. The tachyon profile on the inner radius can depend the  modulus $a$. The simplest possibility is to take  the same tachyon profile on the two boundaries. This case  was  considered in \cite{ar}, see also \cite{all}. 

Another choice was made in \cite{berkeley} where invariance under the interchange of the ends of the of strings
$z\rightarrow a/\bar{z}$ was enforced by tachyon profile $T(X)=u/aX^2$  on the inner boundary.  The role of extra $a$ here is to cancel the factor $\sqrt{h}$ for the inner boundary. We can  recover the tree level contribution by  shrinking  the inner radius of the annulus to  zero and so that the annulus becomes a disk. For the tachyon profile considered  in   \cite{berkeley}  one is still left with  a nonzero tachyon profile at the origin after the radius of the inner circle vanishes. This  corresponds to  inserting  a vertex operator   $<X^2(0)>$ at the origin. There is no doubt  that this  contributes to
the tachyon $\beta$-function. In the supersymmetric case the operator insertion at the origin would  break the  world sheet SUSY  which of course would be  disastrous. Furthermore, in the theory with a symmetric tachyon profile  it is impossible to implement Fishler-Susskind (FS) mechanism to cancel the tadpole divergence. 

A different tachyon profile was used 
in \cite{andreev2}. The tachyon profile was fixed  by imposing   that the partition functions on the annulus and on  the cylinder are the same. As we mentioned earlier it is unclear why it should be true    since 
the theory is not conformal invariant in the presence of a tachyon profile. Furthermore, the analysis in \cite{andreev2} was only carried out for a  constant tachyon profile. If one carries out the calculation   for a  quadratic profile one  finds 
unphysical poles at some values of $u$ in the integration over the modulus. 

From the discussion above it is clear that the boundary terms on the annulus should be of the form
\beg
S_{bndry}=\frac{u}{2\alpha '}\oint_{\rho =1}\frac{d\theta}{2\pi}X^2(\theta)+\frac{au}{2\alpha '}\oint_{\rho =a}\frac{d\theta}{2\pi}X^2(\theta),
\en
\beg
S_{bulk}=\frac{1}{4\pi \alpha '}\int d^2\sigma\sqrt{\gamma}\gamma^{ab}\partial_a X\partial_b X.
\en
Our goal in this paper is to calculate partition function on an annulus  with this boundary interaction. We will follow the methods used for  finding the partition function on the disk. We  start by determining the   Green function
 $G(\theta,\theta ')=<X(\theta)X(\theta ')>$  with both points lying on the same boundary. As on the disk \cite{larsen} we  expand  field $X$ around its classical solution $X_{cl}$,
$X=X_{cl}+\epsilon $ with boundary condition  $\epsilon =0$ on the boundary. The $u$ dependent part of the action is given by $S[X_{cl}]$. Similar  approach was used in \cite{wung}. To find the Green function 
one has to integrate over all possible boundary 
conditions with a $X(\theta )X(\theta ')$ insertion. Details of the calculations are presented in Appendix A. 

From Appendix A we find
that  Green functions are given by
\beg
G(\theta,0)|_{\rho =a}=2\alpha '\sum^\infty_{n=1}{\cal A}_n(u,a)\cos n\theta ,
\en
\beg
G(\theta,0)|_{\rho =b}=2\alpha '\sum^\infty_{n=1}{\cal B}_n(u,a)\cos n\theta ,
\en
where
\beg
{\cal A}_n(u,a)=\frac{nR_n(a)+u}{n^2+u(1+a)nR_n(a)+u^2a},
\en
\beg
{\cal B}_n(u,a)=\frac{nR_n(a)+ua}{n^2+u(1+a)nR_n(a)+u^2a}.
\en

Then partition function  $Z(u)$  can be written in terms of the Green function  as follows
\beg
(2\alpha ')\frac{d\ln Z(u)}{du}=-G(0)|_{\rho =1}-aG(0)|_{\rho =a}.
\en
Inserting the mode expansion  of the Green function we find
\beg
\frac{d\ln Z(u)}{du}=-a {\cal A}_0(u,a)-{\cal B}_0(u,a)-\sum_{n=1}^\infty (a {\cal A}_n(u,a)+{\cal B}_n(u,a)) ,
\en
where coefficients ${\cal A}_n$'s and ${\cal B}_n$'s are given above.  It is easy to see
that
\beg
\label{zeromode}
{\cal A}_0(u,a)=\frac{1}{2}\lim_{n\rightarrow 0}{\cal A}_n(u,a)\hspace{2cm}{\cal B}_0(u,a)=\frac{1}{2}\lim_{n\rightarrow 0}{\cal B}_n(u,a) .
\en
We will use this limit later in our calculations..

The partition function can be cast into the form
\begin{eqnarray}
\label{1}
\ln Z(u)&=&-\frac{1}{2}\ln u-\frac{1}{2}\ln (1+a-ua\ln a)+\sum^\infty_{n=1}\Large [-\ln n^2+\ln (1+a^{2n})\Large ]-\\ 
&&-\sum^\infty_{n=1}\Large [\ln(1+u/n)+\ln (1+au/n)+\ln (1-a^{2n}g_n(u,a))\Large ] ,\label{2}
\end{eqnarray}
where
\beg
g_n(u,a)=\frac{1-u/n}{1+u/n}~\frac{1-au/n}{1+au/n}.
\en
The expression is quite formal and there are several divergences that we have to regularize. The partition  contains several $\ln$-terms.  The first two $\ln$'s arise   from   zero modes. The first one comes 
from the constant mode that controls behavior of the partition function for  small $u$. As expected it is equal to the  tree level  contribution.  There is an additional zero mode 
due  to  $\ln\rho$ function on the annulus  but it  gives only a subleading term. The first $\ln$ in the sum in (\ref{1}) can be reqularized  by  
the standard zeta function regularization
\beg
\prod^\infty_{n=0}\frac{1}{n^2}=\frac{1}{2\pi}.
\en
Below  we will see  that this  regularized  result plays an essential role in relative normalization of different sectors in the supersymmetric case. 

The total partition function $Z_{total}=Z_A(a)Z(u)$ is a product of the $u$-dependent part $Z(u)$ and  the partition function $Z_A(a)$ for the annulus with no boundary interaction.  The second $\ln $ in the sum in (\ref{1}) cancels the corresponding part in $Z_A(a)$. The first two $\ln$'s in (\ref{2}) are 
identical to tree level contributions and therefore give a factor proportional to the $Z_D(ua)Z_D(u)$, a product of two disk partition functions.  As for the disk case one has to regularize 
the part of the sum that is proportional to
$u$. This is a source of a nonzero $\beta$-function for the bosonic case. For superstrings the infinite contribution from  the bosons is cancelled by the similar contribution from the fermions leaving 
the $\beta$-function zero as expected. We will consider these terms in more details when we study  the supersymmetric case. The last $\ln$ in  (\ref{2}) comes from  the
oscillators. It is finite unless $a\neq 1$. Correspondingly, the partition function integrated over the modulus $a$ diverges at $a=1$. 
The problem is how to calculate this divergence. The modular 
properties of the partition functions in the presence of a tachyon profile are broken . It is hard to find the behavior of $Z(u)$ when $a\rightarrow 1$ by doing a modular transformation $\tau\rightarrow -1/\tau$. Again the analysis simplifies considerably when we study the supersymmetric case. Therefore we postpone detailed analysis for later sections.

\section{Superstrings}

Having analyzed the bosonic case we now turn to the supersymmetric tachyon profile.
We start with type IIA theory with a single non-BPS D9-brane and consider  a  tachyon profile   only in one spatial direction. This   corresponds to the  decay of the D9-brane to a BPS D8-brane. The more general case is obtained by a straight forward generalization.
The supersymmetric version of the boundary interaction is given by
\cite{moore},\cite{harvey}
\beg
\label{bndry1}
S_{bndry}=\oint\sqrt{h}\frac{d\theta}{2\pi}\left (\frac{1}{2\alpha'}T^2(X)-\eta\dot{\eta}-\psi^\mu\eta\partial_\mu T(X)\right ),
\en
where the tachyon on both boundaries is of the form $T(X)=\sqrt{u}X$. On each boundary we also  have an auxiliary fermion $\eta (\theta)$. This is the  same fermion that was introduced in \cite{wittenK}
for IIB and in \cite{horava} for IIA theory.  For zero tachyon profile, the $\nsf$ sector is absent since  non-BPS D-branes do not couple to  R-R-fields. 

Next let us turn to the fermionic zero modes.  They appear when  the fermion  $\eta$ is periodic on the boundary. If the  interaction term of the form $\psi^\mu\eta\partial_\mu T(X)$ is absent  the integration over this zero mode results in a vanishing  functional integral. On the other hand if the interaction term is present the functional integral does not vanish. For antiperiodic  fermions $\eta$  the functional
integral with no  tachyon profile  is $\sqrt{2}$. This factor of  $\sqrt{2}$ is precisely the one  that makes difference between non-BPS and BPS D-brane tensions. 

To integrate 
 out the  auxiliary fermions we rewrite the boundary interaction in a non-local form\cite{moore}  
\beg
S_{bulk}=\frac{1}{4\pi}\int d^2z\left (\frac{2}{\alpha '}\partial X\bar{\partial} X+\Psi\bar{\partial}\Psi + \bar{\Psi}\partial\bar{\Psi}\right ),
\en
\beg
\label{bndry}
S_{bndry}=u\oint_{\rho =1}\frac{d\theta}{2\pi}\left (\frac{1}{2\alpha '}X^2(\theta)+\psi_b (\theta)\frac{1}{\partial_\theta}\psi_b (\theta)\right )+
ua\oint_{\rho =a}\frac{d\theta}{2\pi}\left (\frac{1}{2\alpha '}X^2(\theta)+\psi_a (\theta)\frac{1}{\partial_\theta}\psi_a (\theta)\right ),
\en
where $\psi_a$ and $\psi_b$ are boundary fermions 
\beg
\psi_{a,b}=i(\sqrt{iz}\Psi\pm i\sqrt{-i\bar{z}}\bar{\Psi}).
\en
The square roots arise from the change of coordinate from $z$ to $\theta$ on the boundary.
The sign in the above expression depends on the sector considered. 

We define the inverse derivative
\beg
\frac{1}{\partial_\theta}\psi (\theta)\equiv \frac{1}{2}\int_0^{2\pi} d\theta\epsilon (\theta -\theta ')\psi (\theta ')
\en
with $\epsilon (x)=1$ for $x>0$ and $\epsilon (x)=-1$ for $x<0$.

It turns out that in order to determine the partition function we do not need to know the precise form of boundary interaction (\ref{bndry}).
The world sheet supersymmetry is quite restrictive. When the boundary interaction is turned off, the ${\rm {(-1)}^F R}$ sector  contribution to the partition function vanishes. 
It is reasonable to expect  that this is still the case when  the boundary interaction is turned on since it does not break world sheet supersymmetry. In fact this is what happens on  the disk. For periodic fermions the expectation
value $<S_{bndry}>$ vanishes.  
On each boundary the Green function in the ${\rm {(-1)}^FR}$ sector is given by
$$G_f(\theta ,\theta ')\equiv <\psi\frac{1}{\partial_\theta}\psi (\theta)>=-<X^2(\theta)/(2\alpha ')>.$$
It is easy to see that
the function $G_f$ is periodic as it should since we have inserted a factor ${\rm {(-1)}^F}$. 
In the open string sector the angle variable 
 $\theta$ corresponds to time.
By a simple change of moding from integer to half integer one can go  to  the R sector  in open string channel or to the $\nsf-{\rm NS}$ sector  in the closed string channel. It is clear that there are  
no zero mode contributions and the Green function on the inner boundary takes the form
\beg
G_f^{\rm R}(\theta, 0)|_{\rho =a}=-\sum^\infty_{s=1/2}{\cal A}_s (u,a)\cos s\theta .
\en
To obtain the Green function on the outer boundary we have to change 
the coefficient  ${\cal A}$ to the coefficient  ${\cal B}$.

To obtain the Green function in the NS sector we have to find a way to transform from the R sector to the NS sector. It is easy to see that if we shift the time in the closed string sector by 1 it amounts to  switching from R sector to NS sector. This shift  can be accomplished by  the modular transformation $a\rightarrow -a$. When performing this modular transformation one has to be careful not to change $a$ that comes from $\sqrt{h}$ factor in the boundary interaction. Carrying out the modular transformation we find
\beg
G_f^{\rm NS}(\theta, 0)|_{\rho =a}=-\sum^\infty_{s=1/2}{\cal A'}_s (u,-a)\cos s\theta ,
\en   
where prime in ${\cal A'}$ means that the combination $au$ does not change sign.  
Again to obtain the Green function on the outer boundary one has to replace  ${\cal A}$ with  ${\cal B}$. 
We still need to consider 
$\nsf$ sector which is absent when $u=0$. The Green function  $G_f$ in this sector is obtained 
by  changing the  moding from half integer to integers.  Carrying out this change of moding we find
\beg
G_f^{\nsf}(\theta, 0)|_{\rho =a}=-\sum^\infty_{n=1}{\cal A'}_n (u,-a)\cos n\theta .
\en 
To find the zero mode contribution 
we write the fermionic part of the boundary interaction in the form
\beg
S_{bndry}=u\oint_{\rho =1}\frac{d\theta}{2\pi}\left (\psi_b (\theta)\frac{1}{\partial_\theta}\psi_b (\theta)+...\right )+
v\oint_{\rho =a}\frac{d\theta}{2\pi}\left (\psi_a (\theta)\frac{1}{\partial_\theta}\psi_a (\theta)+...\right ) . 
\en
Here $v$ is an arbitrary coupling on the inner radius of the annulus. If we set   $v=a u$ it reduces to the boundary interaction studied in this paper. Using the property (\ref{zeromode}) one finds
\beg
{\cal A}_0=-\frac{1}{2v}\hspace{2cm}{\cal B}_0=-\frac{1}{2u},
\en
which gives 
\beg
\frac{1}{Z}\frac{dZ}{du}=\frac{1}{2u}+...\hspace{2cm}\frac{1}{Z}\frac{dZ}{dv}=\frac{1}{2v}+...
\en
After a simpler  integration we have 
\beg
Z_{\nsf}(u,a)\sim \sqrt{u v}=u\sqrt{a}.
\en

Note the result is proportional to $\sqrt{a}$.
Therefore the partition function $Z_\nsf$ vanishes when $a  \rightarrow 0$. This is exactly what one would expect since in the limit $a  \rightarrow 0$ the annulus reduces to a disk and on the disk the partition function $Z_\nsf$ vanishes. There is another reason why the factor  $\sqrt{a}$ is important.  For large $u$ it 
will cancel a similar factor arising from bosonic zero modes to give the correct partition function for a BPS D-brane.

When determining the partition function it is convenient to  split it into two parts: the zero mode part and the oscillator part. The  contribution from the oscillators is proportional
to the usual partition function for the annulus $Z_A(a)$ . This  is just a normalization constant in the integration over $u$. 
The partition function for the disk without the zero mode contribution is given by
\beg
Z_D(u)\equiv \frac{1}{\sqrt{2}}4^uu\frac{\Gamma^2 (u)}{\Gamma (2u)}.
\en
This function has two important limits. When $u=0$, we find  $Z_D(0)=\sqrt{2}$ which is just the  contribution from the integral over the boundary fermion $\eta$.
For large $u$, we have
$Z_D(u) \sim \sqrt{2 \pi u}$.

Let consider the Ramond sector. The sums appearing are now integer moded rather than half integer moded
\beg
\frac{1}{Z}\frac{dZ}{du}=-(a{\cal A}_0+{\cal B}_0)-\sum_{n=1}^\infty (a {\cal A}_n(u,a)+{\cal B}_n(u,a))+\sum_{s=1/2}^\infty (a {\cal A}_s(u,a)+{\cal B}_s(u,a)).
\en
Following the analysis on the   disk\cite{moore} we find
\beg
\frac{1}{Z}\frac{dZ}{du}=-(a{\cal A}_0+{\cal B}_0)-2\sum_{n=1}^\infty (a {\cal A}_n(u,a)+{\cal B}_n(u,a))+\sum_{n=1}^\infty (a {\cal A}_{n/2}(u,a)+{\cal B}_{n/2}(u,a)).
\en
The zero mode contribution is easy to integrate. It gives the factor
\beg
\frac{1} {\sqrt{u(1+a-ua\ln a)}}.
\en
The oscillator  part can be cast into the form 
\beg
\label{ZR}
Z\sim Z_D(u)Z_D(ua)\left ( \prod^\infty_{n=1}\frac{1}{n^2}\right ) \frac{\prod^{\infty}_{n=1}(1-a^{2n})}{\prod^{\infty}_{s=1/2}(1-a^{2s})}  
\frac{\prod^{\infty}_{n=1/2}(1-a^{2s}g_s(u,a))}{\prod^{\infty}_{n=1}(1-a^{2n}g_n(u,a))}.
\en
Again $\zeta$-function regularization gives a factor $1/(2\pi)$ instead of first infinite product. In the  NS sector  the factor $ (1-a^{2s}g_s(u,a))$  in (\ref{ZR}) is replaced by  $ (1+a^{2s}g_s(u,a))$ . The part that is proportional to the  product of the partition functions on the disk remains the same as in the Ramond sector. The zero 
mode part is also the same since it comes only from bosonic strings. The expression for the partition function in the $\nsf$-sector is quite different. First of all there is a fermionic contribution proportional to
the product of the disk partition function that exactly cancels similar contribution arising  from the bosonic sector. The second difference is that the factor $1/(2\pi)$ from the bosonic part is  cancelled by 
the same factor from the fermionic part. Finally  we have a contribution from   fermionic zero modes. 

Let summarize our results. 
For NS and R sectors the  partition function is a product
\beg
Z(u,a)=Z^{(0)}(u,a)Z_D(u)Z_D(ua)f^7(0,a)f(u,a),
\en
where $Z^{(0)}$ is the zero mode contribution 
\beg
Z^{(0)}_{NS,R}(u,a)=\frac{1}{2\pi}\frac{1} {\sqrt{u(1+a-ua\ln a)}},
\en
and 
\beg
f^{R,NS}(u,a)=\frac{1}{a^{1/8}}\frac{\prod^{\infty}_{n=1/2}(1\pm a^{2s}g_s(u,a))}{\prod^{\infty}_{n=1}(1-a^{2n}g_n(u,a))}.
\en
The  plus sign for NS sector and  minus  sign is for R sector.
For $\nsf$ sector the partition function is given by
\beg
Z(u,a)=Z^{(0)}(u,a)f^7(0,a)f(u,a),
\en
where the zero mode contribution is
\beg
Z^{(0)}(u,a)=\frac{\sqrt{ua}}{\sqrt{1+a-ua\ln a}}
\en
and 
\beg
f^{\nsf}(u,a)=\sqrt{2}\frac{\prod^{\infty}_{n=1}(1+a^{2n}g_n(u,a))}{\prod^{\infty}_{n=1}(1-a^{2n}g_n(u,a))}.
\en
The factor $\sqrt{2}$ is a normalization factor that is need in order for the function  $f(0,a)$ to reduce to the corresponding function  with a vanishing  tachyon profile. To obtain the full partition function we need 
to sum over different sectors and integrate over the modulus $a$
\beg
Z_{total}=\int^1_0\frac{da}{a}({\rm NS}(u,a)-\nsf(u,a)-{\rm R}(u,a)).
\en
Let us consider two limits of this partition function. First one is  $u=0$. In this limit  the contribution from $\nsf$ sector vanishes and  we are  left with 
\beg
Z_{total}=\frac{1}{\pi\sqrt{u}}\int^1_0\frac{da}{a\sqrt{1+a}}Z_A(a),
\en
where $Z_A$ is the partition function for non-BPS D9-branes with no tachyon profile. Note the factor $1/\sqrt{1+a}$ in the integration measure for the partition function. It has a very simple explanation.  We can rewrite the boundary action in the form 
\beg
F^2+\eta\dot{\eta}+T(X)F+\psi\eta\partial T(X).
\en
If we eliminate the auxiliary
 field $F$ the boundary action  reduces to   (\ref{bndry1}). When the tachyon profile  vanishes the boundary interaction has the form 
$F^2+\eta\dot{\eta}$.  For vanishing $u$ the integration over the  bosonic $F$ contributes the factor $1/\sqrt{1+a}$ to  the partition function.

For large  $u$ we find that the partition function is given by
\beg
\label{ZtotalBPS}
Z_{total}=\int^1_0\frac{da}{a\sqrt{-\ln a}}Z_A^{BPS}(a),
\en
where $Z_A^{BPS}(a)$ is the partition function of a BPS D-brane. This is a  formal expression since the partition for  BPS D-branes vanishes.
The factor  $1/\sqrt{-\ln a}$ appearing in eq. (\ref{ZtotalBPS}) is cancelled  by a similar factor arising from the integration  over the momentum. This can be seen by writing  $a= e^{-\pi/t}$, where $t$ is the open string time.  For large $u$ we indeed have a single D8-brane. For finite value of $u$ a more careful analysis is required. There are two issue on has to deal with. First the tachyon field gives rise to a divergence that must be reqularized. The second problem is
that there is a tadpole divergence. 
The usual way to deal with tachyon divergence is to  analytically  continue. In this case the partition function and  the tachyon potential  become complex. 
One can think of this  as a signal of instability of the system. The tadpole divergence is more crucial. Logarithmic divergences unlike power ones are physical.
They could possibly give a contribution to the tachyon $\beta$-function or to $\beta$-functions of other  fields. These issues will be analyzed in the next section in more details.

\section{Tadpole divergence.}   

Having constructed the partition function we are ready to study tadpole divergences. First let us consider a constant tachyon profile
 $T=c$. The tachyon potential is
\beg
\label{pot}
V(c^2)\sim Z(c^2)=e^{-c^2}\int^1_{1/\Lambda}\frac{da}{a}e^{-c^2a}Z_A(a).
\en
Above we have introduced  a cutoff $\Lambda$ in order to regularize our potential . The infinite part of the integral is $16~e^{-c^2}\ln (\Lambda/c^2)$. Clearly, for  finite  $c$  we need to 
introduce a counterterm to cancel this divergence.  On the other hand if one takes limit $\Lambda\rightarrow\infty$ in such a way that $\Lambda/c^2$ remains finite the divergence disappears 
 and there is no need to introduce  counterterms. This shows that  the theory has two different phases. In the first phase  with finite $c$  a counterterm has to be introduced to cancel the infinity. This counterterm  will contribute to the  to the graviton $\beta$-function. It is important to note that although there are  divergences the $\beta$-function for the tachyon still vanishes.  In the second phase  with $\Lambda/c^2$  finite the theory is finite. For a constant tachyon profile
one cannot get   a BPS D-brane since R-R-sector is  missing. Tadpole divergence is usually cancelled by the  Fischler-Susskind (FS) mechanism\cite{FS}, see also\cite{SUSYFS}. The insertion
of the dilaton vertex operator at zero momentum $g_o2\ln \Lambda\Psi^\mu(0){\bar{\Psi}}_\mu(0)$ at the origin of the disk  makes the total partition function finite even for a 
 constant tachyon profile.  The coupling constant $g_o$ is the open string coupling constant. 

Next we will consider the linear tachyon profile. For small $u$ the partition function has a $1/\sqrt{u}$
behavior. This arises from  a potential that has a  $T^2$  dependence after one ingrates over the bosonic field X.  The tachyon potential is determined by a constant tachyon profile.

We are now ready to analyze 
 the tadpole divergence for the linear profile. Again we have  to introduce a cutoff $\Lambda$  when we integrate  over the modulus $a$. We find that there are two phases depending on the order we take the limits. In the first phase  we let $ a \rightarrow 0 $ while $u$ is fixed. 
The second phase is obtained by first taking the limit $u  \rightarrow\infty$ and then integrating over the modulus $a$.  Since in the end one has to take  $\Lambda \rightarrow \infty$ limit  the second phase corresponds to  BPS D-brane and no 
counterterm is needed. Let us now  consider the first phase. We keep $u$ finite while  $a\rightarrow 0$ . Since the contribution of R-R sector is proportional to $\sqrt{a}$
there is no divergence in this sector as $a\rightarrow 0$. The only divergence comes from NS and ${\rm {(-1)}^F NS}$ sectors. The closed string tachyon is 
cancelled between these two  sectors but the contribution from the massless state still remains. We find that the logarithmic divergence has the form
\beg
\label{log}
\ln \Lambda\frac{\sqrt{2}}{2\pi\sqrt{u}}Z_D(u)2\left ( 7+\frac{1-2u}{1+2u}\right ).
\en
 
The $\sqrt{2}$ factor comes from $Z_D(au)$ at $a=0$.  The factor 2 arises because  NS and R sectors give identical contributions. The numerical factor 7 in the bracket
is the contribution from directions with vanishing tachyon profile whereas the last factor is the contribution from the direction with a non-vanishing tachyon profile. Since we are using the FS mechanism to cancel divergences arising from the boundary of moduli space we need to add an operator at the origin of a disk.  

To determine this contribution we need to find  
the  Green function in the bulk of the disk in presence of the tachyon profile $T(X)=\sqrt{u}X$ . A simple calculation gives  
\beg
<\Psi (z)\bar\Psi (\bar w)>=\frac{1}{\sqrt{z\bar w}}\left (-\frac{\sqrt{z\bar w}}{1-z\bar w}+2u\sum^\infty_{s=1/2}\frac{1}{s+u}{(z\bar w)}^s \right ).
\en
At  the origin it takes the form
\beg
<\Psi (0)\bar\Psi (0)>=-\frac{1-2u}{1+2u}.
\en 
This shows 
 the tadpole  can be cancelled by a  dilaton vertex operator  
\beg
g_0\frac{2}{\pi}\ln\Lambda e^{-\phi} e^{-\tilde\phi}\Psi_\mu\bar\Psi^\mu
\en
at zero momentum at the origin. Here $\phi$ and $\tilde\phi$ are bosonized superghosts.
Note that the vertex operator is in $(-1,-1)$ picture. 
 This shows that the
FS mechanism works for the linear tachyon profile.

One would like to transform the above vertex operator to  $(0,0)$ picture. When conformal invariance is not broken
the dilaton vertex operator  in $(0,0)$ picture is
 $\partial X^\mu\bar\partial X_\mu$ integrated over the disk. Any point in the interior of the disk can be brought
 to the origin by a conformal transformation. Then the integration over the disk becomes an integration over the
 volume of the conformal group. 
The quadratic tachyon profile on the boundary breaks conformal invariance. Given a vertex operator at a fixed 
point on the disk there is no simple way to find a corresponding operator in $(0,0)$ picture. For related
discussion see \cite{emil}.    
At the fixed points of RG flow where the conformal invariance is restored it is again possible to write the vertex operator as an integral over the disk in (0,0) picture.

Let us summarize what has been done so far. If  the large $\Lambda$ limit is taken first while $u$ is kept finite
we are in  the first phase. In this phase the NS-NS divergence can be eliminated by introducing the dilaton
vertex operator on the disk. If then let $u$ approach  infinity R-R anomaly appears. 
This indicates that
the system is  entering  the second phase. In the second phase the large $u$ limit is taken first.  The total 
partition function now is just the partition function of a D8-brane. The tadpole is cancelled by R-R anomaly. 
No dilaton vertex operator is needed.
This phase transition happens at $u\sim\Lambda$.

It is easy to see that if the tachyon profile on the inner boundary were as in \cite{berkeley} ($\sqrt{h}$ factor is absent) this divergence would be proportional to 
$Z_D^2(u)$ which would be impossible to cancel by FS mechanism.

So far we have considered space filling D9-branes that do not have any Dirichlet directions. The discussion for 
lower dimensional D-branes can proceed along the same lines except one subtle point.  Each Dirichlet direction kills
one zero mode for the corresponding boson and thus adds the factor $1/\sqrt{-\ln a}$. Denote $d$ as a number  Dirichlet directions. The contribution of massless modes is proportional to 
\beg
\int_{1/\Lambda}\frac{da}{a}{(-\ln a)}^{-d/2}.
\en
For $d=0,1,2$ there is a tadpole and the dilaton vertex operator must be inserted on the disk. For all other
lower dimensional D-branes the integral converges in the closed string limit. Therefore the phase transition
occurs only for D9, D8 and D7 branes. 

Although the factor $1/\sqrt{-\ln a}$ also affects behavior of  the integrand in the open string limit $a\rightarrow 1$ the main divergence in this case comes from the open string tachyon. We consider this divergence in the next section.  
 
\section{The fate of the open string tachyon}

The open string sector of the theory is determined by the behavior of function $f(u,a)$ in the limit $a\rightarrow 1$. To find this limit we can not use modular properties of the function $f(u,a)$ since we do not know how it transforms 
under  the modular 
transformation $\tau\rightarrow-1/\tau$.  In Appendix B we show how to determine leading behavior of the function
$f(u,a)$ in the limit $a\rightarrow 1$.

To study the limit $a \rightarrow 1$ while $u$ is constant  we will consider different  sectors separately. We start 
with the NS sector  and find 
\beg
\label{ns}
Z\sim e^{-\pi ut/3}e^{\pi^2u/3}\frac{\sinh 2\pi u}{\sinh \pi u}\frac{1}{2\pi}\frac{1}{\sqrt{2u}}\frac{1}{2}2 e^{\pi t}+O(1/t),
\en
where the factor of 2 is needed in order to have the  correct $u\rightarrow 0$  limit.  
In the  $\nsf$ sector we find 
\beg
\label{nsf} 
Z\sim e^{-\pi ut/3}e^{\pi^2u/3}\sinh \pi u\frac{1}{\sqrt{2u}}\frac{1}{\pi}e^{\pi t}+O(1/t)
\en
and in the  R sector we have 
\beg
\label{r}
Z\sim e^{2\pi ut/3}e^{-2\pi^2 u/3}\frac{1}{\sqrt{2u}}\frac{1}{2\pi}2+O(1/t), 
\en
where again the factor of 2 was inserted to have the correct $u \rightarrow 0$ limit. 
As we mentioned in Appendix B this analysis only works for the supersymmetric case. In the bosonic case one encounters divergences that we do not know how to analyze.
In \cite{ar},\cite{berkeley} this behavior was investigated by calculating the central charge of strings on the cylinder. But as the conformal invariance is broken
the partition functions for the cylinder and for the annulus do not coincide to each other. They are the same only in the two conformal limits. Therefore it is not 
clear whether the central charge of the theory on the cylinder determines the leading behavior of the partition function on the annulus in the open string limit.

The most interesting conclusion one can derive from the asymptotic  behavior considered above  is that  the tachyon disappears when 
$u=3$ in NS and $\nsf$ sectors. The ground state becomes massive if $u>3$. 
In the limit of $u \rightarrow \infty$ the contribution from  NS and $\nsf$ sectors are equal  and cancel each other as they contribute with an opposite sign.
There is a tachyon in the R sector for any non zero $u$ but its contribution is multiplied by a factor that decreases with $u$.

Since there is a tachyon in the open string sector the integral over modulus $a$ is divergent and has to be regularized.
If we try to follow the analysis of the close string sector  by introducing a cutoff, we find for the open string has power law divergences. These are clearly unphysical. 

There is a way to treat  this divergence. One can  analytically  continue the partition function. The price one has to pay for this is that  the partition function as well as
the action for the tachyon becomes  complex. This is a common situation in the field theory when a loop corrected effective potential with a  negative second derivative is
considered. Since the mass of the tachyon is related to the second derivative of the potential it is very likely that the loop corrected action for the tachyon is also
complex. The imaginary part of the action displays the fact that the system is not stable and decays\cite{berkeley}. The imaginary part  is proportional to the decay rate\cite{andreev}.
Following \cite{marcus} the imaginary part of the tachyon action is 
\beg
\label{Im}
{\rm Im}\int^\infty_R\frac{dt}{t}t^{-\beta}e^{bt}=\frac{\pi}{\Gamma (1+\alpha)}b^\beta .
\en
Since there is  a tachyon in the R sector for any finite $u$ one has to check that in the large $u$ limit the imaginary part of the partition function goes to zero so 
one would have a stable configuration. It is indeed true since the imaginary part in the R sector is of the form $u^{10}e^{-2\pi^2u/3}/\sqrt{u}$ as $u$ is large.

If there are some Dirichlet directions the formula (\ref{Im}) is still useful. Since $1/\sqrt{-\ln a}=\sqrt{t/\pi}$
the dimension of the brane affects only the constant $\beta$ in (\ref{Im}). 

\section{Tachyon potential and tachyon action}

In this section  we study the tachyon potential $V(T^2)$. For large values of the tachyon field \mbox{ $T$} the potential has the following behavior
\beg
V(T^2)\sim e^{-T^2}\left (1+Ag_o\int^\infty_{T^2/\Lambda}\frac{da}{a}e^{-a}\right ),
\en
where the first term is the contribution from the disk  and the second term is the contribution  from  the annulus. The coupling constant $g_o$ is the open string coupling constant  and $A$ is some constant. As before we find two phases. The first one is obtained when $T^2/\Lambda\rightarrow 0$ and $T$ is large but finite. In this limit the potential has the form
\beg
V(T^2)\sim e^{-T^2}\left(1+Ag_o\left(-\gamma-\ln \frac{T^2}{\mu}+ ...\right)\right),
\en
where $\gamma$ is  the Euler constant. Above we have added a counter term to subtract off the infinities following our  discussion in the previous chapter. The constant 
 $\mu$ is a finite scale that appears when we regulate the theory and subtract the infinite contribution. The sign of the potential depends on the ratio of $T^2/\mu$ and it can be  either negative or positive.  

In the second phase $T^2/\Lambda\rightarrow\infty$ as $T^2 \rightarrow\infty$ . In this limit the potential has the leading behavior of the form
\beg
V(T^2)\sim e^{-T^2}\left (1+Ag_oe^{-T^2/\Lambda}\left (\frac{\Lambda}{T^2}+...\right ) \right ).
\en
As we discussed before no counter term is required in this limit. In both phases the potential vanished when $u \rightarrow \infty$

In ref.  \cite{moore} it was shown that the linear tachyon profile gives the correct ratio for different D-brane tensions. Following their analysis we will study the decay of a single D9-brane to a BPS D8-brane. In order to get ratio of the D-brane tensions we have to study  tachyon action for both small and large $u$. 
In the large  $u$ limit one obtaines the BPS D8-brane.
 The D8  tension  does not receive any loop corrections as the one loop partition function vanishes for large $u$ 

The D9-brane tension is obtained from the $u\rightarrow 0$ limit.
The loop correction  to the D9-brane tension does not vanish. Since the $\nsf$ sector  does not contribute  to the non BPS D-brane tension, we will not consider it below. Before proceeding with our calculation it is useful to collect some of our results. The potential receives a contribution at tree level $V^{(0)}(T^2) $ and  at one loop $V^{(0)}(T^2)$.  The tree level  contribution is given by 
\beg
V^{(0)}(T^2)=e^{-T^2}
\en
and the one loop contribution is given by
\beg
V^{(1)}(T^2)=\frac{1}{2}\int^1_0\frac{da}{a}e^{-(1+a)T^2}({\rm NS}(0,a)-\rm{R}(0,a)).
\en
The tachyon action has the form
\beg
S=S_0\left (\frac{1}{\sqrt{2\pi}}\frac{1}{\sqrt{u}}Z_D(u)+g_o\frac{1}{2}\frac{1}{2\pi}\frac{1}{\sqrt{u}}Z_D(u)Z_{NS,R}(u)\right ),
\en
where 
\beg
Z_{NS,R}(u,a)=\int^1_0\frac{da}{a}\frac{1}{\sqrt{1+a-ua\ln a}}Z_D(au)({\rm NS}(u,a)-\rm{R}(u,a))
\en
and $S_0$ is a  constant proportional to the D9-brane tension. 

The D-brane tension can be determined from the non-derivative terms of the effective action. These are  controlled by the zero mode structure. On the disk this amounts to replacing the bosonic zero mode factor $\frac{2}{\sqrt{u}}$   by the following expression \cite{moore} 
\beg
\sqrt{\frac{2}{u}}=\int^\infty_{-\infty}\frac{dX}{{(\alpha '\pi)}^{1/2}} e^{-u/(2\alpha ')X^2}.
\en
For the contribution due to the annulus we have a similar expression
\beg
\frac{1}{2}\frac{2}{\sqrt{u}}=N \int^\infty_{-\infty}\frac{dX}{\sqrt{\alpha '\pi}}V^{(1)}(T^2)=\frac{1}{2} N \int^\infty_{-\infty}\frac{dX}{\sqrt{\alpha '\pi}}
\int^1_0\frac{da}{a}e^{-(1+a)u/(2\alpha ')X^2}({\rm NS}(0,a)-\rm{R}(0,a)).
\en
The normalization constant $N$ is determined as follows
\beg
N^{-1}=\frac{1}{\sqrt{2}}\int^1_0\frac{da}{a}Z_{NS,R}(0).
\en
Inserting these expressions to the tachyon action we find
\beg
S=\frac{S_0}{{(\pi\alpha ')}^5}\int d^{10}x\left (
\frac{1}{2\pi\sqrt{2}}V^{(0)}(T^2) Z_D(2\alpha '\dot{T})+g_o\frac{1}{2}\frac{N}{2\pi}V^{(1)}(T^2)Z_D(2\alpha '\dot{T})Z_{NS,R}(2\alpha '\dot{T})
\right ).
\en

The factor $\sqrt{2}$ from each $Z_D$ was incorporated in the potential. That is the origin of 
$1/\sqrt{2}$ in the tree level part (one auxiliary fermion $\eta$) and $1/2$ in the one loop part (two auxiliary fermions). Since the zero mode normalization is fixed 
by the disk action the extra factor $\sqrt{2}$ in $N$ is needed. Again the contribution from $\nsf$ sector of the theory is omitted since its presence has no affect  
to the D9-brane tension. Finally one can look at the behavior of the action in small $u$ limit leaving only the part proportional to the potential
\beg
S\sim \frac{S_0}{{(\pi\alpha ')}^5}\left ( \frac{1}{\sqrt{2\pi}}\int d^{10}xV^{(0)}(T^2)+g_o\sqrt{2}\frac{1}{2\pi}\int d^{10}x V^{(1)}(T^2)\right )  .         
\en
The tachyon action must be of the form
\beg
S=\int d^{10}x (T_9^{(0)}V^{(0)}+g_oT_9^{(1)}V^{(1)}).
\en
Comparing it with our expression for the action it is easy to find that
\beg
\frac{T_9^{(0)}}{T_9^{(1)}}=\sqrt{\pi},
\en
which has a clear explanation. All one loop behavior is accumulated in the potential $V^{(1)}(T^2)$. The only difference from the disk case is factors $1/\sqrt{2\pi}$ and
$\sqrt{2}$. The first comes from the extra set of boundary modes for the annulus, the second is from the extra auxiliary fermion compare to the disk case. 
To get an expression for the loop corrected D-brane tension one has to set $T=0$
\beg
T_9=T_9^{(0)}\left ( 1+\frac{1}{\sqrt{\pi}}g_o\int^1_0\frac{da}{a}({\rm NS}(0,a)-\rm{R}(0,a))\right ),
\en 
which is as expected proportional to the partition function of non BPS D9-branes\cite{lambert}.

\section{Multiple D-branes}

So far we have considered tachyon profiles with a single D-brane.  In this section we will generalize our discussion by including several unstable D-branes. The tachyon profile for these configurations will be represented in terms of a hermitian matrix. The supersymmetric boundary interaction involves several boundary fermions $\eta^I$'s. The index $I$ labels the Chan-Patton (CP) factor and for each CP-factor we have a boundary fermion. 

Let us concentrate on the  R-R sector. The boundary fermions play an important role in determining whether the R-R sector is present.
To see how this works let us first consider a single D-brane with a tachyon profile  $T_a(X)=y_a X$ on the inner boundary and $T_b(X)=y_b X$ on the outer boundary. In the R-R sector the boundary fermions $\eta_{a,b}$ are periodic and hence have zero modes. For a vanishing tachyon profile  the functional integral over the boundary fermions vanishes because of the zero modes. When the   tachyon profile is  non-vanishing  the  interaction term $\psi\eta\partial T$  will soak up the zero modes. The functional integral over the boundary fermions gives a term proportional to $y_ay_b$. 
Next let us consider a more complicated example of two non BPS D9 branes  in the type IIB theory.
These D9 branes can decay into a single D7 brane that then can decay into a BPS D6-brane. 
In this case  we have three CP factors and the tachyon profile has the form \cite{moore}
\beg
T(X)=\sum_{I=1}^3 y_IX^I\gamma^I=\sum_{I=1}^3 T^I(X)\gamma^I,
\en 
where $\gamma^I$'s are hermitian and  they satisfy a Clifford algebra $\{\gamma^I,\gamma^J\}=2\delta^{IJ}$.
The boundary interaction has now the form \cite{larsen}
\beg
\label{bndryCP}
-\eta^I\dot{\eta}^I-\psi^\mu\partial_\mu T_I(X)\eta^I.
\en 
On the annulus we have two sets of $\eta^I$'s, one for each boundary.
In order to get a non-zero result when integrating   over  boundary fermions  the product of the  $\eta^I_{a,b}$'s. has to appear.
The functional integral will then give a term proportional to $y_1^2y_2^2y_3^2$. From this it is obvious
that the partition function  in  the R-R sector vanishes if one of the f $y_I$'s is zero.

If we set $y_3=0$ and take $y_1$ and $y_2$ to 
infinity  this describes the decay of a D9 brane to a non-BPS D7 brane.
The final configuration is non BPS so the R-R sector is absent and there is no coupling to the R-R fields.

The non BPS D-branes are very similar to D-${\rm \bar{D}}$ systems. In this case tachyon is complex. Consider tachyon condensation in a 
system of D8-${\rm \bar{D}8}$ branes in type IIA.  The system decays to a single D6-brane. For this system we have two CP factors and the tachyon profile is given by
\beg
T(X)=
\left (
\begin{array}{cc}
0&\sqrt{u}X^9+i\sqrt{v}X^8\\
\sqrt{u}X^9-i\sqrt{v}X^8&0
\end{array}
\right )
=T_1(X)\sigma_1+T_2(X)\sigma_2 .
\en

There are  no  off diagonal entries in this tachyon profile. The boundary interaction is almost the same as in (\ref{bndryCP}) . The only difference is that instead of three boundary fermions we now have only two boundary fermions $\eta^I$'s on  each boundary.
A BPS D6 brane is obtained in  the limit  $u \rightarrow \infty$ and $v\rightarrow \infty$. The R-R sector is non-vanishing and the D6 brane couples to the R-R fields.
If we set $v=0$ and let $u \rightarrow \infty$, the R-R sector is absent and the final state describes a non-BPS D7-brane.

The discussion can be easily generalized to any number of D-branes and anti D-branes. The only difference is that  the particular form of 
the boundary interaction (\ref{bndryCP}) is  different\cite{larsen}.

\section{Tachyon condensation for codimension four}

In this section we will revisit the phase transition  we encountered previously, this time   in codimension four.
Recall that the first phase occurs when $u$ is small compared to the cutoff $\Lambda$. In this phase 
divergences are cancelled  by a dilaton vertex operator  at the origin of the disk amplitude.
In the R-R sector the partition  diverges when  $u \rightarrow \infty$. This infinity is know as R-R anomaly.
For a BPS state the R-R anomaly is cancelled by a tadpole divergence in the NS-NS sector.
In the first phase  tadpole divergences are  cancelled by the FS mechanism.
One can see that there is a phase transition since  for large $u$  the two point function $<\Psi\bar\Psi> $ is non-vanishing for a single D-brane .

In the second phase we  first let $u\rightarrow\infty$ for finite $\Lambda$. In this phase there are no divergences and hence no need for FS mechanism. In this case we have an almost BPS state and the NS-NS and R-R divergences cancel each other.

The dilaton vertex operator gives a non-zero contribution in the large $u$ limit, except in codimension four. To see this let us consider a tachyon profile in p spatial directions. The  
NS-NS divergence is proportional to
\beg
8-p+\sum^p_{i=1}\frac{1-2u_i}{1+2u_i}.
\en 
For large $u_i$'s and $p=4 $ we have
\beg
<\Psi^\mu (0)\bar{\Psi}_\mu (0)>=0.
\en
This means that  for large $u_i$'s we approach a  BPS state with no contribution from the dilaton operator.

Next let us consider an example with $p=4$.  In the type IIA theory  2(D8-${\rm \bar{D}8}$) system will decay  into a  stable  BPS D4-brane.  We have a dilaton vertex operator at the origin. As  we take the limit $u_i\rightarrow \infty$ in all four directions the contribution from the dilaton vanishes and there is no phase transition. 

Similar statements hold true for a decay of 2(D9-${\rm \bar{D}9}$) and 2(D7-${\rm \bar{D}7}$) in type IIB. 
As mentioned above D-branes with number of Dirichlet directions greater then two do not have tadpole divergence.

\section{Conclusion}

In this paper we have analyzed boundary string field theory at the one loop level. We have argued that  the tachyon profiles on the inner and the  outer circles of the annulus must be the same. We  find that this is consistent with large $u$ limit and the FS mechanism. 
We have analyzed tadpole divergences in superstring theory and showed that for a linear tachyon profile the tadpole divergence is cancelled by the FS mechanism. 
For D9, D8 and D7 branes we find two different phases. One phase has a finite scale and an insertion of the dilaton operator at the origin of the disk amplitude. This phase is an almost  non-BPS phase. The other phase is an almost BPS phase. The phase does not have any scale or any divergences and no counterterms are required. For branes 
of codimension bigger then 2 there is no tadpole divergence and there is no phase transition. The tachyon condensation for D9, D8 and D7 branes can also be viewed as a smooth process if the number of dimensions being reduced is equal to four. 

We have studied the  open string limit of the annulus partition function. If the tachyon profile is not zero
the partition function for the annulus does not coincide with the partition function for the cylinder, and one cannot use modular properties to obtain the open string limit. Nevertheless, it is possible to find the 
leading behavior of the superstring partition function in this limit.   
For finite value of  $u$ the open string tachyon  disappears  in the NS and $\nsf$ sectors. On the other hand  we find  an open string tachyon at any non-zero value of  $u$ in the   R sector.  Associated with the tachyon there is a divergence
when $a$ is close  to the boundary of modula space $a\sim  1$. One can treat this divergence by analytic continuation.   The price to be paid is 
that the action for the tachyon becomes complex. 

We find that the loop corrected tachyon potential has a logarithmic term. This might have  applications to cosmology of the rolling tachyon since Wick's rotation of the time direction does not change the form of the potential\cite{sen}. 
We also comment on how the boundary fermions $\eta^I$'s switch off and on the R-R part of the partition function. 

Our calculations have shown that tachyon condensation at one loop depends on how many Dirichlet directions are  present and how many dimensions are  reduced. These differences are related to the one loop
divergences and therefore cannot be seen at tree level. It would be interesting to see  if these can give  rise to a new "loop corrected`` K-theory? 

\section{Acknowledgments}
This work was supported in part by funds provided by the DOE under grant DE-FG03-84ER40168

\appendix
\section{Calculation of the bosonic Green function}
 
The most general harmonic function on the annulus is
\beg
X_{cl}(\rho , \theta )=X_0+k\ln\rho + \sqrt{\frac{\alpha '}{2}}\sum^\infty_{n=1}(\rho^n(C_n e^{in\theta}+C_{-n} e^{-in\theta})+
\rho^{-n}(D_n e^{in\theta}+D_{-n} e^{-in\theta})).
\en
The reality condition gives
  $C_n^*=C_{-n}$, $D_n^*=D_{-n}$.
On the boundary
harmonic function is given by 
\beg
X_{cl}(\rho =a)=X_a+\sqrt{\frac{\alpha '}{2}}\sum^\infty_{n=1}(A_n e^{in\theta}+A_{-n}e^{-in\theta}),
\en
\beg
X_{cl}(\rho =1)=X_b+\sqrt{\frac{\alpha '}{2}}\sum^\infty_{n=1}(B_n e^{in\theta}+B_{-n}e^{-in\theta}) 
\en
with $A_n^{*}=A_{-n}$, $B^{*}_n=B_{-n}$. Then the relations between $X_{cl}$ in the bulk and its values on the boundaries are
\beg
C_{\pm n}=\frac{B_{\pm n} -A_{\pm n} a^n}{1-a^{2n}},
\en
\beg
D_{\pm n}={a^n}\frac{A_{\pm n}-a^n B_{\pm}}{1-a^{2n}},
\en
\beg
k=-\frac{X_b-X_a}{\ln a},
\en
\beg
X_0=X_b .
\en 
The action can be determined  in terms of boundary modes $A_{\pm n}$'s and $B_{\pm n}$'s. A simple  calculation gives
\begin{eqnarray*}
S[X_{cl}]&=&-\frac{{(X_b-X_a)}^2}{2\alpha '\ln a}+\frac{u}{2\alpha '}(X_b^2+aX_a^2)+\\
&&+\frac{1}{2}\sum^\infty_{n=1}\left [(n R_n(a)+u)B_nB_{-n}+(nR_n(a)+au)A_nA_{-n}\right ]-\\
&&-\frac{1}{2}\sum^\infty_{n=1}nK_n(a)(B_nA_{-n}+A_nB_{-n}),
\end{eqnarray*}
where
\beg
R_n(a)=\frac{1+a^{2n}}{1-a^{2n}},
\en
\beg
K_n(a)=\frac{a^n}{1-a^{2n}}.
\en
The last step is to integrate over the boundary modes and find the Green function on the boundary
\beg
G(\theta ,0)=\int\prod^\infty_{n=1}dA_ndA_{-n}dB_ndB_{-n}e^{-S[X_{cl}]}X_{cl}(\theta)X_{cl}(0){\Big/}
\int\prod^\infty_{n=1}dA_ndA_{-n}dB_ndB_{-n}e^{-S[X_{cl}]}.
\en
Consider  contribution from  zero modes $X_a$ and $X_b$
\beg
\frac{1}{2\alpha '}G_{0}(\rho = 1)\equiv {\cal B}_0(u,a)+...=\frac{1}{2u}\left (\frac{1-ua\ln a}{1+a-ua\ln a }\right )+...  
\en
\beg
\frac{1}{2\alpha '}G_{0}(\rho = a)\equiv {\cal A}_0(u,a)+...=\frac{1}{2u}\left (\frac{1-u\ln a }{1+a-ua\ln a}\right )+... 
\en
The zero mode contribution controls the behavior of the partition function for  small values of $u$. To find the full Green  function one has to integrate over all other
boundary modes. With zero modes omitted the Green function is 
\beg
G(\theta,0)|_{\rho =a}=2\alpha '\sum^\infty_{n=1}{\cal A}_n(u,a)\cos n\theta ,
\en
\beg
G(\theta,0)|_{\rho =b}=2\alpha '\sum^\infty_{n=1}{\cal B}_n(u,a)\cos n\theta ,
\en
where
\beg
{\cal A}_n(u,a)=\frac{nR_n(a)+u}{n^2+u(1+a)nR_n(a)+u^2a},
\en
\beg
{\cal B}_n(u,a)=\frac{nR_n(a)+ua}{n^2+u(1+a)nR_n(a)+u^2a}.
\en
We can generalize this Green function to the case where we have  different boundary conditions on the two boundaries
\beg
{\cal A}_n(u,a)=\frac{nR_n(a)+u}{n^2+(u+v)nR_n(a)+uv},
\en
\beg
{\cal B}_n(u,a)=\frac{nR_n(a)+v}{n^2+(u+v)nR_n(a)+uv}.
\en
The parameter $u$ corresponds to the outer boundary and the parameter v corresponds to the inner boundary.
When we have the same tachyon profile on the two boundaries $v=a u$.

\section{Open string limit for functions $\ln f(u,a)$}
Let us start with the R sector.
We want to find the leading behavior  of the following function 
\begin{eqnarray}
\label{NSF}
\sum^\infty_{s=1/2}\ln\left (1-a^{2s}\frac{1-u/s}{1+u/s}\frac{1-au/s}{1+au/s}\right )&-&\sum^\infty_{n=1}\ln\left (1-a^{2n}\frac{1-u/n}{1+u/n}\frac{1-au/n}{1+au/n}\right )-\\
-\sum^\infty_{s=1/2}\ln (1-a^{2s})&+&\sum^\infty_{n=1}\ln (1-a^{2n}) 
\end{eqnarray}
in the limit  $a\rightarrow 1$.
The sum over half integers can be written in more convenient form $\sum_s b_s=-\sum_n b_n+\sum_n b_{n/2}$. The next step is to expand $\ln$'s in a power series. After some algebra we find
\begin{eqnarray}
\label{uff}
-\sum_{k=1}^\infty\sum_{n=1}^\infty\frac{1}{k}
\bigg [a^{nk}{\left (\frac{1-2u/n}{1+2u/n}\frac{1-2au/n}{1+2au/n}\right )}^k&-
&2a^{2nk}{\left (\frac{1-u/n}{1+u/n}\frac{1-au/n}{1+au/n}\right )}^k\bigg ]+\\
\label{uff2}
+\sum_{k=1}^\infty\sum_{n=1}^\infty\frac{1}{k}(a^{nk}&-&2a^{2nk}).
\end{eqnarray}
Let us try to expand  the sum (\ref{uff2}) in powers of $u/n$. The zeroth order term of this expansion is  cancelled by $\ln$'s expansion from the second line of (\ref{uff2}). The second order term is
proportional to $1/n^2$. The sum $\sum_n a^{nk}/n^2$ is convergent at $a=1$. The higher order terms have even better convergence property. Therefore the only divergence in the sum at 
$a=1$   comes from the linear  $u/n$ term. The sum is of the form
\beg
4u(1+a)\sum_{n=1}^\infty\frac{1}{n}\left (\frac{1}{1-a^n}-\frac{1}{1-a^{2n}}\right ).
\en
When $a$ is close to 1 one can write $a$ as $1-\epsilon$. Since $a=e^{-\pi/t}$ where $t$ is the open string time, $\epsilon$ corresponds to $\pi/t$. We are interested in  
divergent and constant parts. Extracting these we find
\beg
4u\sum_{n=1}^\infty\frac{1}{n^2}\left ( \frac{1}{\epsilon}-1\right)=4u\zeta (2)\left (\frac{1}{\epsilon}-1\right ).
\en
To find the finite part we set $a=1$ in (\ref{NSF}) and throw away the divergent contribution leaving us  with the  finite part 
\beg
\sum_{n=1}^\infty (-2\ln (1+2u/n)+4\ln(1+u/n))=-\ln Z^2_D(u).
\en
This last step is not quite rigorous since the right hand side contains linear in $u$ terms but this is the natural regularization of the sum on the left hand side.
In the  R sector the leading contribution is
\beg
-\ln Z_D(u)+4u\frac{\pi^2}{6}\left (\frac{1}{\epsilon}-1\right ),
\en
where we have used $\zeta (2)=\pi^2/6$.
The same procedure for NS sector gives the leading behavior
\beg
-\ln Z^2_D (u)+\sum_{n=1}^\infty\ln (1+{(u/n)}^2)-2u\frac{\pi^2}{6}\left (\frac{1}{\epsilon}-1\right ).
\en
In the $\nsf$-sector we find
\beg
\sum_{n=1}^\infty\ln \frac{1+{(2u/n)}^2}{1+{(u/n)}^2}-2u\frac{\pi^2}{6}\left (\frac{1}{\epsilon}-1\right ).
\en 
To express the infinite products that appear in the above formulas one needs 
\beg
\prod^\infty_{n=1}(1+{(u/n)}^2)=\frac{\sinh \pi u}{\pi u}.
\en
Performing  the calculations  outlined above it is possible to loose some constant factors that do not depend on either $\epsilon$ or $u$. To fix these terms  one has to look at the small $u$ behavior
which is easy to analyze in the open string limit.
  
We want to stress a feature that enables us to calculate the leading behavior. The terms of order $O(\epsilon^0)$ are finite, they still can be summed to $\zeta (2)$. In the case
of bosonic strings this is not true. If one is looking at $O(\epsilon^0)$ terms there is a sum of the form $\sum_n 1/n$. For  superstring the  sums from bosons and 
fermions  cancel against each other.

\end{document}